# Post-deposition annealing and interfacial ALD buffer layers of Sb$_2$Se$_3$/CdS stacks for reduced interface recombination and increased open-circuit voltages

Thomas Paul Weiss[1], Ignacio Minguez-Bacho[4], Elena Zuccalà[1], Michele Melchiorre[1], Nathalie Valle[2], Brahime El Adib[2], Tadahiro Yokosawa[3], Erdmann Spiecker[3], Julien Bachmann[4,5], Phillip J. Dale[1], Susanne Siebentritt[1]

[1] University of Luxembourg, Department of Physics and Materials Science, 41, rue du Brill, L-4422 Belvaux, Luxembourg

[2] Luxembourg Institute of Science and Technology, Materials Research and Technology, 41, rue du Brill, L-4422 Belvaux, Luxembourg

[3] Friedrich-Alexander University Erlangen-Nürnberg, Institute of Micro- and Nanostructure Research, and Center for Nanoanalysis and Electron Microscopy (CENEM), IZNF, Cauerstraße 3, Erlangen, 91058 Germany

[4] Friedrich-Alexander University of Erlangen-Nurnberg, Chemistry of Thin Film Materials, IZNF, Cauerstraße 3, 91058 Erlangen, Germany

[5] National Centre for Nano Fabrication and Characterization, Oersteds Plads, Building 347, 2800 Kongens Lyngby, Denmark

## Abstract

**Currently, Sb$_2$Se$_3$ thin films receive considerable research interest as a solar cell absorber material. When completed into a device stack, the major bottleneck for further device improvement is the open circuit voltage, which is the focus of the work presented here. Polycrystalline thin film Sb$_2$Se$_3$ absorbers and solar cells are prepared in substrate configuration and the dominant recombination path is studied using photoluminescence spectroscopy and temperature dependent current-voltage characteristics. It is found that a post-deposition annealing after the CdS buffer layer deposition can**

**effectively remove interface recombination since the activation energy of the dominant recombination path becomes equal to the bandgap of the $Sb_2Se_3$ absorber. The increased activation energy is accompanied by an increased photoluminescence yield, i.e. reduced non-radiative recombination. Finished $Sb_2Se_3$ solar cell devices reach open circuit voltages as high as 485 mV. Contrarily, the short-circuit current density of these devices is limiting the efficiency after the post-deposition annealing. It is shown that atomic layer deposited intermediate buffer layers such as $TiO_2$ or $Sb_2S_3$ can pave the way for overcoming this limitation.**

# 1 Introduction

Thin film photovoltaic technologies offer great opportunities for electrical power generation, such as light weight and flexible modules, low energy payback time and a low $CO_2$ equivalent footprint [1, 2]. $Sb_2Se_3$ contains only abundant and low-toxic elements, can be used as an absorber layer and recently received considerable attention as a photovoltaic material used both in superstrate and substrate device design configurations. The bandgap can be engineered from 0.9 eV to 1.7 eV using Bi [3] and S [4] alloying and thus is ideally suited for single- [5] and tandem-junction [6, 7] solar cells. Recently, record efficiencies of 9.2 % for $Sb_2Se_3$ [8] and 10.5 % for $Sb_2(Se,S)_3$ [9] solar cells have been achieved. As pointed out by Chen and Tang, the open circuit voltage is currently limiting device performance [10]. Figure 1 depicts the status of $Sb_2Se_3$ based solar cells, i.e. without the addition of S, for substrate and superstrate architectures, which are briefly reviewed in the following. The underlying data is listed in Appendix A. The efficiency for superstrate devices shows a clear correlation with the open circuit voltage (Figure 1a) due to good charge carrier collection (Figure 1b), i.e. short-circuit current densities above 25 mAcm$^{-2}$. In contrast, the short-circuit current densities are rather low ($\lesssim$ 25 mAcm$^{-2}$) for substrate devices, even though this architecture yields the highest open circuit voltages. The fill factors are similar for both types (Figure 1c). Considering both, $V_{OC}$ and $J_{SC}$, recent improvements in efficiency are rather driven by increasing $J_{SC}$ (Figure 1d). In fact, the highest efficiencies are obtained in substrate configuration for rough micro-rod absorber morphologies [8, 11, 12], which facilitates charge-carrier extraction, however, imposes additional non-radiative recombination losses due to an increased junction surface area.

Here, the voltage limitation for $Sb_2Se_3$ in substrate configuration is studied. Open circuit losses can severely be influenced by the quality of the $Sb_2Se_3$ front interface. In this work, we use two approaches to improve the $Sb_2Se_3$ front interface: a) post-deposition annealing and b) ultrathin interfacial layers by atomic layer deposition (ALD). Recently, a post-deposition annealing after the buffer layer deposition led to a record open circuit voltage of 506 mV [13]. We show that such a post-deposition annealing can effectively improve the interface as well as the bulk material quality of the $Sb_2Se_3$/CdS stack. While previous studies in superstrate configuration pointed out that CdS is not suitable as a buffer layer, it is shown here that in substrate configuration the $Sb_2Se_3$/CdS interface does not impose a major problem such as a pinned Fermi level or a conduction band cliff. $Sb_2Se_3$/CdS stacks as well as finished solar cell devices post-deposition annealed in air prepared in this study reached open circuit voltages as high as 485 mV and thus demonstrates the validity and effectiveness of such a treatment. In addition, we investigate the use of amorphous interfacial $TiO_2$ and $Sb_2S_3$ buffer layers, to improve the interface quality. ALD provides conformal coatings of uniform and continuous layers over nanostructured surfaces. In previous works, it has been demonstrated that the use of ultrathin interfacial layers via ALD in $Sb_2S_3$ and $Sb_2Se_3$ devices reduces interface recombination [8, 14, 15]. As will be shown here, the introduction of these ultrathin layers in the $Sb_2Se_3$/CdS interface results in an enhancement of all the photovoltaic parameters.

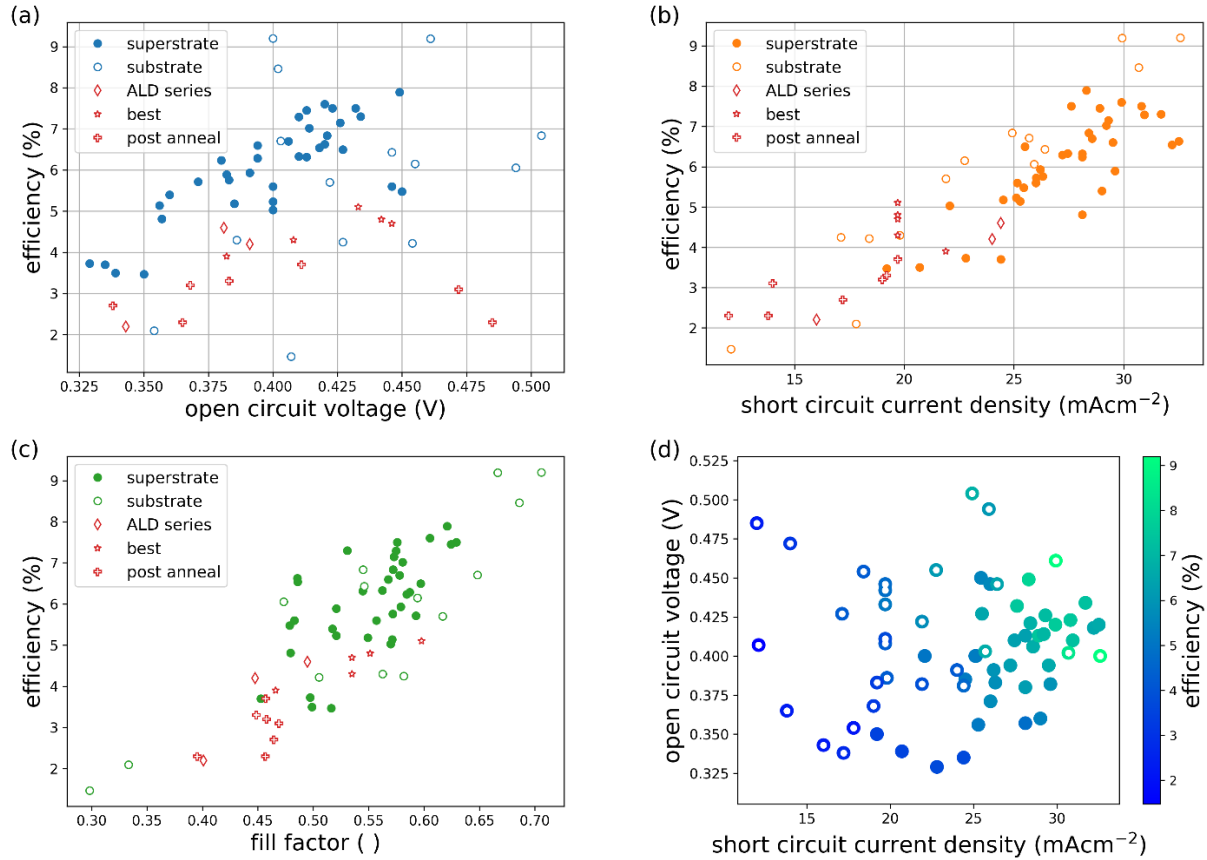

Figure 1 – IV parameters extracted from reported Sb$_2$Se$_3$ based solar cells (see Table 1 in Appendix A) in superstrate (full symbols) and substrate (open symbols) architecture. Data points from solar cells from the current study are added as open red symbols.

## 2 Experimental

The experimental procedure for the fabrication and characterization of Sb$_2$Se$_3$ absorbers and solar cells is depicted in Figure 2 and consists of the following steps. Metallic Sb precursors are thermally evaporated on a soda lime glass/Mo substrate at substrate temperatures below 30 °C. Sb$_2$Se$_3$ thin film absorber layers are subsequently formed by selenization in a 1-zone furnace, where Se powder is added as a Se source. The temperature profile of the selenization process presents three steps: i) heating to approximately 110 °C and flushing with N$_2$ to remove residual oxygen and water from the chamber and the sample surface, ii) selenization at 320 °C (for study with interfacial ALD layer, described below) or 380 °C (for study with post-deposition annealing, described below) for 30 min with a 1.2 bar N$_2$ background pressure, iii) post-treatment at 140 °C for 120 min. During the first 60 min of the post-treatment, no changes to the pressure/annealing atmosphere are undertaken. After 60 min of the post-treatment, the chamber is pumped down below 10$^{-3}$ mbar to remove any elemental Se, which might have adsorbed on the sample surface. Afterwards, the sample is cooled down to room temperature. The temperature profiles used in this study are shown in Supplementary Figure 1.

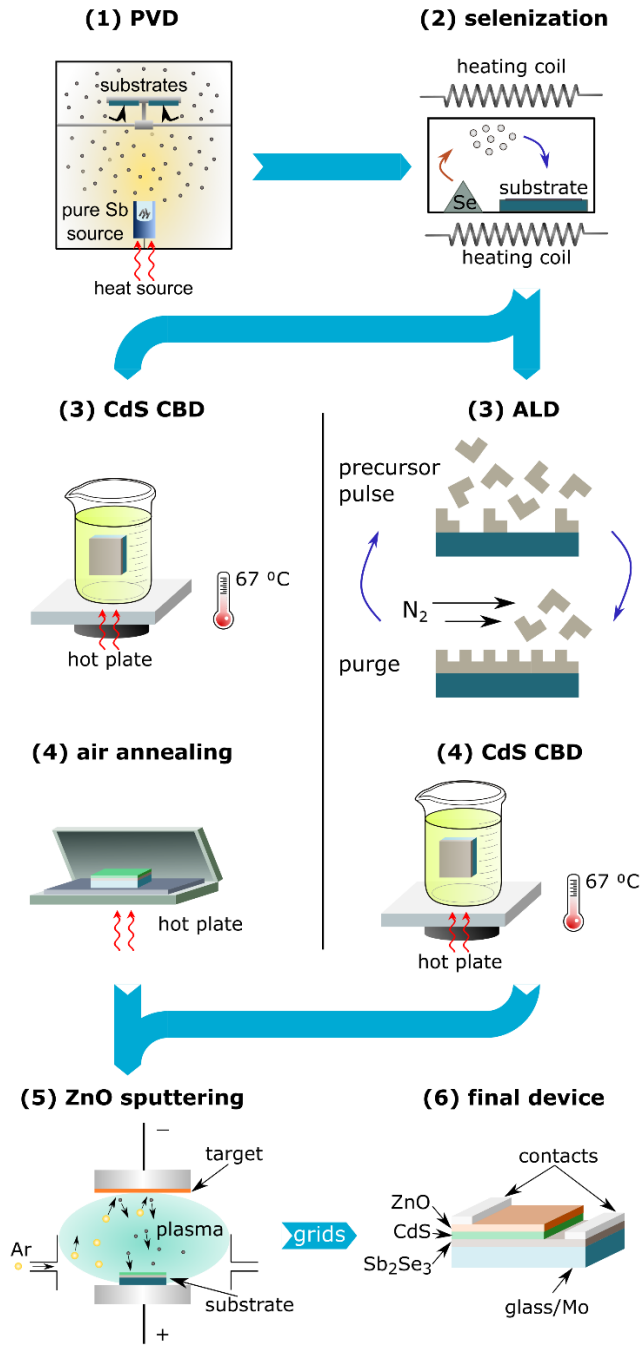

Figure 2 – Experimental procedure for the growth of Sb$_2$Se$_3$ based absorbers and solar cells. PVD stands for phyiscal vapor deposition, CBD for chemical bath deposition and ALD for atomic layer deposition.

Prior to the deposition of the CdS buffer layer, the absorbers were etched in 5 wt.% KCN for 30 s, which generally improves the open circuit voltage of the final device (Supplementary Figure 2a). This improvement is likely due to removal of oxide species, similar to other selenide semiconductors [16]. A CdS buffer layer is deposited on top of the absorber layer by means of chemical bath deposition (CBD) at 67 ºC adapted from ref. [17]. The thickness of the CdS layer is approximately 80 nm and was achieved by two subsequent CdS processes. The thicker CdS layers generally lead to a higher open circuit voltage (Supplementary Figure 2b). The window layer consists of a sputtered ZnO double layer, with 80 nm ZnO and 380 nm ZnO:Al. Ni/Al dot-contacts are deposited by e-beam evaporation. Individual solar cells are scribed mechanically with areas of 0.2 cm$^2$.

Post-deposition annealing of $Sb_2Se_3$ absorbers is performed on a hotplate in air, either after CdS deposition or on full devices. The annealing temperatures range between 100 to 300 °C and the annealing time is fixed to 5 min.

- **Post-deposition annealing after CdS deposition**: two $Sb_2Se_3$/CdS absorber/buffer stacks are prepared from the same precursor run in two nominally identical selenization processes at 380 °C. The CdS buffer layer is deposited prior to the post-deposition annealing to prevent possible degradation, as for instance observed for Cu(In,Ga)Se$_2$ absorbers [16, 18]. After the CdS deposition, both samples are cut into several pieces. One piece of each sample is kept without a post-deposition annealing to serve as reference devices. Both reference devices behave very similar and therefore no further distinction between the two different samples is made. The other pieces are treated by a post-deposition annealing for 5 minutes each at a different temperature between 100 °C and 300 °C. All $Sb_2Se_3$/CdS stacks (reference or post-deposition annealed) are then further characterized as described below and processed into solar cell devices.

- **Post-deposition annealing of a full device:** a full solar cell device is post-deposition annealed successively with the conditions mentioned above starting at 100 °C up to 300 °C in steps of 50 °C. After each post-deposition annealing, current-density voltage and external quantum efficiency characteristics are measured. The precursor for the device originated from a different deposition run as the precursor for the post-deposition annealing study after the CdS deposition. The experimental parameters for the selenization process are the same as for the samples used for the "*Post-deposition annealing after CdS deposition*" series.

For some samples, intermediate thin films of $TiO_2$ or $Sb_2S_3$ are grown using ALD prior to the CdS buffer layer deposition. Absorber layers for this experiment are selenized at 320 °C (see Supplementary Figure 1 for the temperature profile) and no post-deposition annealing was applied.

The interfacial ALD $TiO_2$ layer is deposited using an Arradiance Gemstar XT benchtop reactor. The precursors utilized are Titanium(IV) isopropoxide (TTIP) and ultrapure $H_2O$. The temperature of the TTIP and $H_2O$ are kept at 70 °C and room temperature, respectively. The time sequence for precursor pulse/exposure/purge during the reaction cycles are set to 0.5/40/50 s for both precursors. The sample temperature is kept at 150 °C. The selenized $Sb_2Se_3$ absorbers are coated with 25 $TiO_2$ ALD cycles at a deposition rate of 0.4 Å per cycle.

The interfacial ALD $Sb_2S_3$ is deposited using a homemade hot-wall ALD reactor. The precursors used are tris(dimethylamido)antimony(III) ($Sb(NMe_2)_3$, 99.99%, Sigma-Aldrich), and $H_2S$ (3% vol in $N_2$, Air liquide) and are kept at 40 °C and room temperature, respectively. The precursor pulse/exposure/purge times sequence are 1.5/15/15 s for the Sb-precursor, and 0.2/15/15 s for $H_2S$. The substrate temperature is kept at 120 °C during the deposition. The $Sb_2Se_3$ samples are coated with 10 $Sb_2S_3$ ALD cycles at a deposition rate of 0.5 Å per cycle.

After the interfacial ALD layers and the CdS buffer layer, the absorbers are finished into solar cell devices as described above.

Low energy secondary ion mass spectrometry (SIMS) technique is used to investigate the composition of the thin films and in particular the possible diffusion of elements at the interface. With this aim, SIMS depth profiles are acquired with a 1 keV $Cs^+$ bombardment using a CAMECA SC-ULTRA SIMS instrument. Main elements Cd, S, Se as well as trace elements (O, Na and K) are analyzed from the top surface of CdS to the $Sb_2Se_3$ layer. The sputtering rate is converted into depth by assuming a constant sputtering rate of about 2.7 nm/s through the whole thickness analyzed.

Cross-sections of the specimens for transmission electron microscopy (TEM) are prepared by standard lift-out technique using a ThermoFischer Helios Nanolab 660 dual beam focused ion beam (FIB)/SEM. Scanning TEM (STEM) and high-resolution TEM (HRTEM) studies are performed with a (ThermoFischer) Titan[3] Themis microscope operated at 300 kV. Energy-dispersive X-ray spectroscopy (EDX) combined with STEM (STEM-EDX) are carried out by using highly sensitive Super-X detectors to obtain elemental maps.

Reflectance spectra, $R$, are acquired using a Perkin Elmer 950 UV-Vis-NIR spectrophotometer on SGL/Mo/Sb$_2$Se$_3$/ZnO/Al:ZnO stacks, i.e. prior to the deposition of the Ni/Al front contact. From these measurements, the light entering the solar cell is calculated as $1-R$ and governs the maximal achievable external quantum efficiency.

For photoluminescence (PL) measurements, a red diode laser with a wavelength of 660 nm is focused onto the CdS covered Sb$_2$Se$_3$ absorber layer. Spectral correction is carried out by measuring a white light reference spectrum at the sample position. In this study, two different laser powers are used to acquire the PL spectra: i) 10 mW, which induces only minor heating of the sample and yields a rather constant PL yield over time and ii) 100 mW, which induces a significant amount of heating and results in an increasing PL yield over time. If not stated otherwise, the PL spectra shown in this manuscript are recorded with a laser power of 10 mW.

Current density-voltage (*jV*) characteristics are measured using a class AAA sun simulator in 4-probe configuration under standard test conditions. A Si reference cell is used for intensity calibration.

The external quantum efficiency (EQE) is measured by lock-in amplification using a halogen and xenon lamp as a light source. The light is focused on a spot without grid shading. A Si and InGaAs reference diode are used for calibration. From the EQE the internal quantum efficiency (IQE) is calculated as $IQE = EQE/(1-R)$.

Temperature dependent current-voltage characteristics are recorded in a closed cycle He cryostat with a halogen lamp as a light source. The distance of the lamp to the sample is adjusted to match the same short circuit current density as measured under the sun simulator. A temperature sensor is glued on a soda lime glass substrate next to the sample under test to estimate the temperature of the absorber.

## 3   Results and Discussion

### 3.1   Device analysis

The impact of the post-deposition annealing temperature of the Sb$_2$Se$_3$/Cds stack on the final finished solar cell devices is analyzed by their *jV* characteristics in the dark and under illumination. The corresponding device parameters are shown in Figure 3a-d. The following trends can be observed

i. The open circuit voltage is rather constant for annealing temperatures up to 200 °C, where only a small drop at 150 °C observed. However, for post-deposition annealing temperatures of 250 °C and above, a clear increase in the $V_{OC}$ is observed. These devices will be subject to a more detailed investigation as shown in the following. It is noted that the bandgap does not change (Supplementary Figure 5a).
ii. The short-circuit current density decreases with increasing post-deposition annealing temperature. The highest values are obtained for the reference devices and the device annealed at 100 °C.
iii. The fill factor remains rather constant independent of the post-deposition annealing temperature.
iv. The efficiency decreases as a result of a decreasing $J_{SC}$ with increasing post-deposition annealing temperature.

Figure 3e shows the best jV characteristics. The dark jV characteristics shift towards higher voltages with increasing post-deposition annealing temperature and indicate an improved material quality due to a smaller recombination current. The dark curves also show no significant influence of shunting, as indicated by the flat diode current around 0 V. Fitted values of the shunt resistance yield values in the order of 10 k$\Omega cm^2$ as shown in Supplementary Figure 3a. In contrast, the illuminated curves show a significant positive slope around 0 bias, which indicates voltage dependent carrier collection or an illumination induced shunt path. In order to assess the diode quality, i.e. the recombination current, the IV curves were fitted using the iv-fit routine [19]. The voltage dependent current density $J(V)$ is fitted to a 1-diode model, including a series, $r_s$, and a shunt, $R_{sh}$, resistance, according to

$$J(V) = J_0 \left[ \exp\left(\frac{V - Jr_S}{Ak_BT}\right) - 1 \right] + \frac{V - Jr_S}{R_{sh}} - J_{ph} \quad (1)$$

In equation (1), $J_0$ is the saturation current density, $A$ is the diode ideality factor, $k_B$ is the Boltzmann constant, $T$ is the temperature during the acquisition of the jV curve, and $J_{ph}$ is the photo-generated current density. Supplementary Figure 3b shows the experimental data and the corresponding fits for the best devices of each post-deposition annealing. Evidently, the saturation current density $J_0$ is characteristic for the quality of the diode (see for instance section 2.4.5 in ref. [20]) and directly influences $V_{OC}$ via $V_{OC} = Ak_BT \ln(J_{ph}/J_0)$ (neglecting $R_{sh}$ here). Figure 3f shows the fitted values for $J_0$ and demonstrates a decrease by roughly two orders of magnitude for the device with the highest annealing temperature compared to the reference devices. The smaller value of $J_0$ clearly indicates an improved quality of the diode and thus a reduced recombination current. The parameters for the device with the highest $V_{OC}$ (annealed at 300 °C) are: $V_{OC} = 485$ mV, $J_{SC} = 12.0$ mAcm$^{-2}$, FF = 39.5 % resulting in an efficiency of 2.3 %.

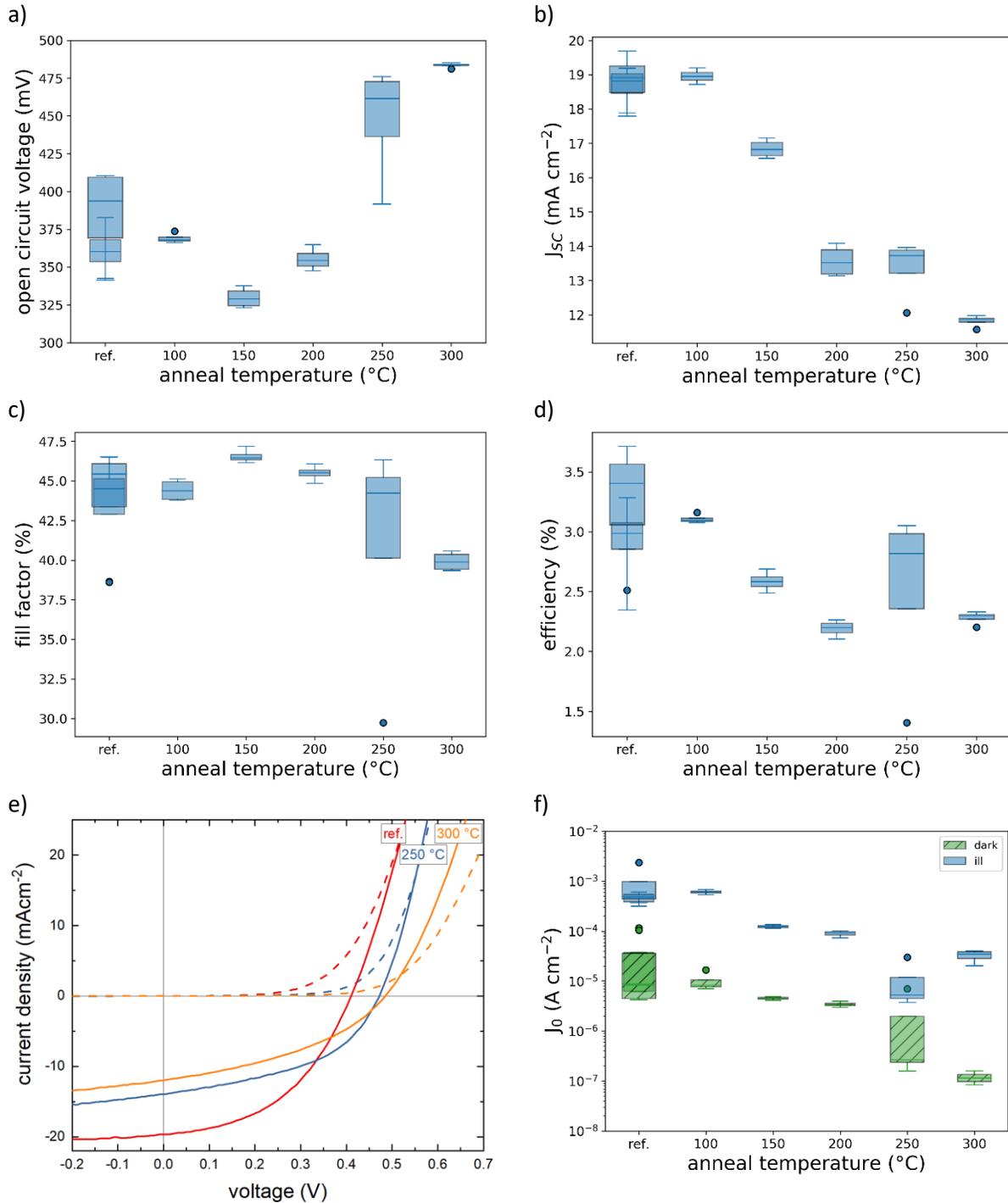

Figure 3 – a-d) Device parameters for solar cells processed with varying post-deposition annealing temperatures after the CdS buffer layer deposition. e) Best jV curves. f) Saturation current density $J_0$ fitted using the 1-diode model (equation (1)).

Mainly due to the decreasing $J_{SC}$ with annealing temperature, the overall efficiency of the devices is reduced. Parasitic resistances can already induce a $J_{SC}$ reduced by a factor of $1/(1 + r_S/R_{sh})$ from $J_{ph}$ (see Supplementary Information section A). However, for the devices presented here, this effect has only a minor influence (Supplementary Figure 4). In order to assess the decrease of $J_{SC}$, Figure 4 shows measurements of the external quantum efficiency. Optical losses due to reflection are indicated by the quantity $(1 - R)$, which indicates the maximal achievable EQE. The as-grown sample provides a maximum charge carrier collection efficiency around 600 nm, which reduces towards longer wavelength (> 700 nm). The collection improves in this long wavelength range by applying a small reverse bias voltage, as can be seen by the plot of $EQE(-0.2V)/EQE(0V)$ on the right ordinate. Due to the application of a reverse bias, the space charge region enlarges and consequently the collection of carriers, which are generated deeper in the Sb$_2$Se$_3$ absorber layer. In contrast, the samples post-deposition annealed at 250 °C and 300 °C show a strongly reduced EQE over the whole wavelength range. The plot of $EQE(-0.2V)/EQE(0V)$ indicates an improved collection at reverse bias over the whole wavelength range. It is noted that the bandgap is not influenced by the post-deposition annealing and yields values around 1.25 eV as determined from the internal quantum efficiency (Supplementary Figure 5a), within the range of bandgaps observed elsewhere (1.18 eV [21] – 1.26 eV [22]). Figure 4b shows the ratio of the IQEs. The increase towards higher wavelength shows that the as-grown device has a better collection over the whole wavelength range, with an increased contribution at longer wavelengths. Such an observation indicates that the collection length decreases upon post-deposition annealing.

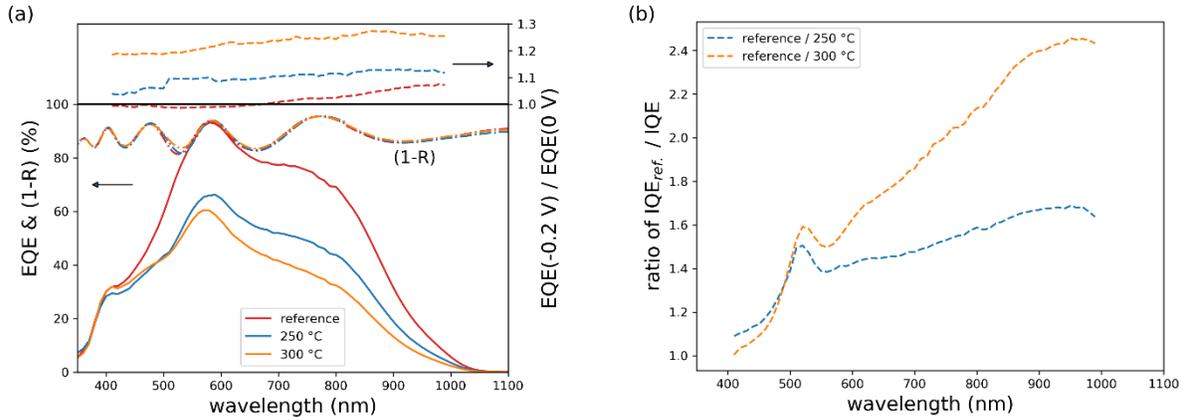

Figure 4 – External quantum efficiency (EQE, solid lines) and reflectance losses (1-R, dash-dotted lines) versus the photon wavelength on the left ordinate. Good collection is achieved in the reference sample around 600 nm and decreased towards longer wavelength. Post-deposition annealed samples at 250 °C and 300 °C show a decreased EQE. The ratio of the EQE recorded at -0.2 V bias and at 0 V bias is shown on the right ordinate. b) The ratio of the reference IQE with the IQE after post-deposition annealing. The as-grown sample shows a better collection over the whole wavelength range, in particular for longer wavelengths.

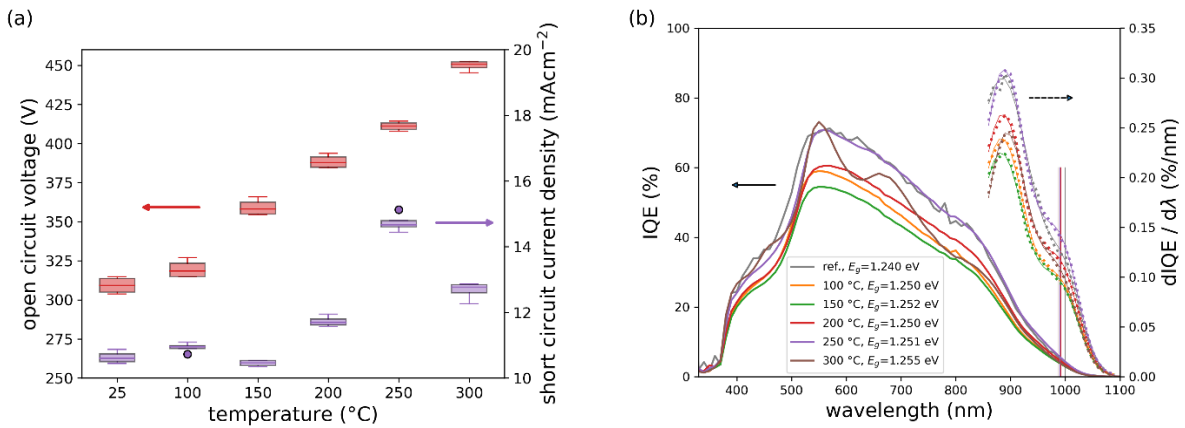

Figure 5 – Open circuit voltage (left ordinate) and short circuit current density (right ordinate) as a function of successive post-deposition annealings (from low to high temperatures) carried out on a finished solar cell device (not on the $Sb_2Se_3$/CdS stacks as shown in Figure 3). The open circuit voltage continues to increase with post-deposition annealing temperature. On the other hand, the short circuit current density exhibits a peak value at 250 °C. The improved short circuit current density is also observed by the IQE measurement (b), which shows an improvement over the full wavelength range. The bandgap does not significantly change upon post-deposition - annealing as indicated by the plot of $dIQE/d\lambda$ (dashed lines, right ordinate).

A similar behavior of an increasing $V_{OC}$ with respect to post-deposition annealing is observed on a complete solar cell (Figure 5). In that case, the post-deposition annealing procedure was carried out successively on a completed solar cell device. Clearly, an increasing open circuit voltage is observed with successive post deposition annealing with increasing temperature. The bandgap of the Sb₂Se₃ does not change as shown in Figure 5b.

In summary we observe an increase in Voc independently of whether we heat Sb2Se3/CdS or the whole stack, and no noticeable change in band gap is observed. To investigate the voltage change further we use PL.

### 3.2 Photoluminescence analysis

Photoluminescence spectroscopy is used to assess the Sb₂Se₃ absorber layer quality in the absence of window layers, which potentially can introduce additional recombination channels. A higher PL yield is expected for reduced non-radiative recombination and hence an improved material quality. Measurements are carried out on the samples listed in **Error! Reference source not found.**, i.e. including the CdS buffer layer.

Normalized PL spectra for the different post-deposition annealing temperature are shown in Figure 6. The energetic peak position is unchanged around 1.20 eV, in agreement with the bandgap values observed from the internal quantum efficiency (Supplementary Figure 5a). The upper left inset shows that the peak FWHM reduces with increasing post-deposition annealing temperature indicating an improved material quality, in terms of band tails and/or homogeneity. The PL yield increases up to a post-deposition annealing temperature of 250 °C in accordance with an improved material quality and the reduction of the FWHM. A decreased PL yield for the 300 °C post-deposition annealed sample is observed with respect to the samples annealed at 200 or 250°C, indicating increased non-radiative recombination, even though this sample has the lowest FWHM. Also, the $V_{OC}$ of this sample is higher. Currently, we cannot explain the lowering of the PL yield at the highest annealing temperature.

An increase of the quasi Fermi level splitting $\Delta\mu$ can be estimated from an increased PL yield according to [23]

$$\Delta\mu = k_B T \ln \frac{Y_{PL,post-annealed}}{Y_{PL,as-grown}} \qquad (2)$$

For the sample annealed at 250 °C, an increased quasi Fermi level splitting of roughly 50 meV is estimated compared to the reference sample, which is in good agreement with the increase of the $V_{OC}$ (within the spread of the data shown in Figure 3a) of the final devices processed from the same samples as used here for the PL characterization. The sample annealed at 200 °C should yield a similar improvement of the open circuit voltage. However, interface recombination dominates this complete solar cell device, which reduces the open circuit voltage (Supplementary Figure 6).

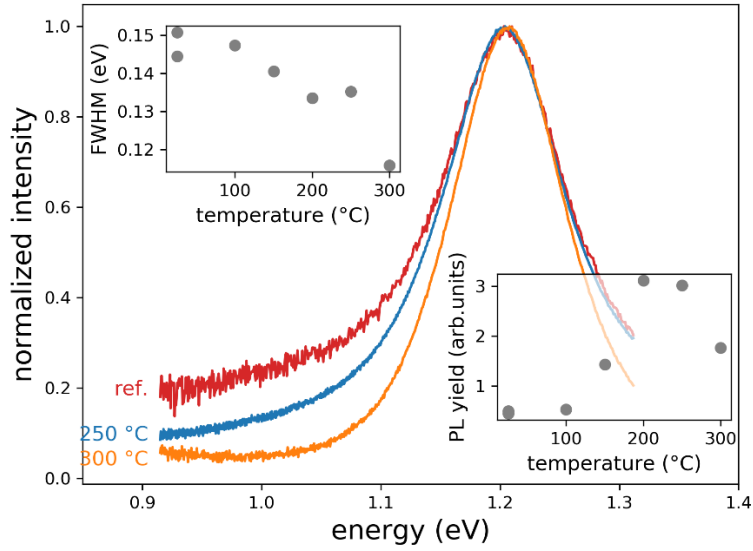

Figure 6 – Normalized photoluminescence spectra of Sb₂Se₃/CdS stacks post-deposition annealed for 5 minutes after the CdS deposition. Excitation power is 10 mW. Upon post-deposition annealing, the FWHM (upper left inset) reduces, whereas the PL yield (lower right inset) increases (with the exception at 300 °C), indicating an improved material quality. No significant energetic shift of the PL peak position is observed.

### 3.3  Dominant recombination path

The saturation current density in equation (1) dictates the onset of the diode current and can be written as [20]

$$J_0 = J_{00} \exp\left(-\frac{E_A}{Ak_B T}\right) \tag{3}$$

with $J_{00}$ the reference saturation current density and $E_A$ the activation energy of the dominant recombination path. For dominating recombination in the space-charge-region or in the quasi-neutral-region an activation energy equal to the bulk bandgap is expected. However, an activation energy below the bulk bandgap indicates dominating interface recombination, which can be caused by a pinned Fermi level, a cliff-like band alignment [24] or a defect layer near the interface [25].

From (1) and (3), the temperature dependent open circuit voltage reads as

$$V_{OC}(T) = \frac{E_A}{q} - \frac{Ak_B T}{q} \ln\left(\frac{J_{SC}}{J_{00}}\right) \qquad (4)$$

where $q$ is the elemental charge. Extrapolation of $V_{OC}(T)$ to 0 K yields the activation energy of the dominant recombination path $E_A$. Figure 7a shows $V_{OC}(T)$ data for the 250 °C post-deposition annealed and its as-grown reference sample as blue circles. The 300 °C post-deposition annealed sample behaves as the 250 °C one (Supplementary Figure 7) and is omitted for clarity. The $V_{OC}$ at room temperature (300 K) is roughly 80 mV higher for the sample air-annealed at 250 °C (see also Figure 3). A similar increase is obtained for the activation energy of $J_0$, which increases from $E_{A,ref} = 1.18 \pm 0.02$ eV to $E_{A,250C} = 1.27 \pm 0.01$ eV, for the as-grown and 250 °C post-deposition annealed sample, respectively. Thus, after post-deposition annealing, the activation energy shifts towards the bulk bandgap value (itself unchanged by post-deposition annealing) as deduced from IQE measurements (Supplementary Figure 5a) and indicates an improved interface quality.

Also, the improved $V_{OC}$ for successive post-deposition annealings of a completed solar cell device (Figure 5) are well explained by an increased activation energy of $J_0$. The as-finished device has a $V_{OC}$ of only 298 mV (at 300 K) with an activation energy of $E_{A,init} = 0.67$ eV, which is well below the bandgap, indicating (near) interface recombination as the main path. However, the device after the final post-deposition annealing step at 300 °C shows an activation energy of $E_{A,fin} = 1.22$ eV, which is close to the bandgap of $Sb_2Se_3$ and explains the improved $V_{OC}$. The bandgap does not change upon the successive post-deposition annealings on the finished device (Supplementary Figure 5b).

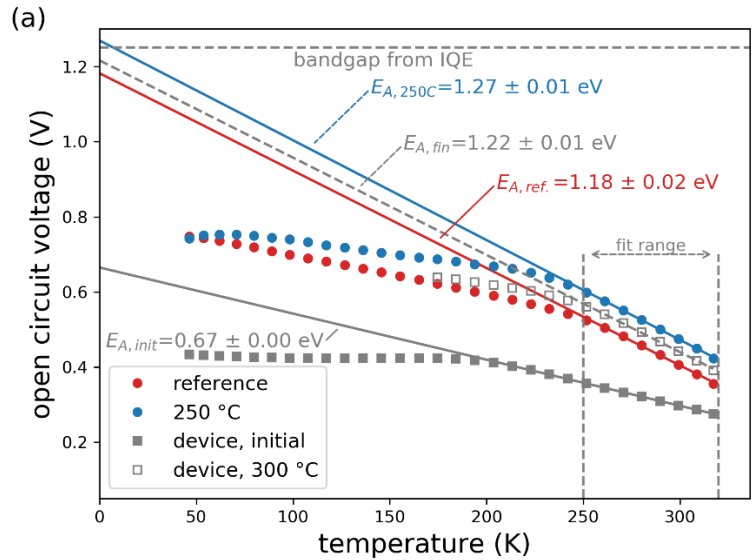

Figure 7 – a) Temperature dependence of the open circuit voltage. Extrapolation of a linear fit to 0 K yields the activation energy $E_A$ of the dominant recombination path. Errors of the activation energy are taken from the fitting confidence interval.

## 3.4 Laser annealing

The devices presented in the previous sections were obtained by post-deposition annealing the Sb$_2$Se$_3$/CdS on a hot plate which involves heating the glass substrate and all the layers simultaneously which is energetically expensive. Previously, it has been demonstrated that laser annealing may be used to improve semiconductor quality without significantly heating the substrate [26] . Here, we investigate whether increasing the laser power by a factor 10 during the PL measurements is sufficiently heating and simultaneously enables the measurement of the PL yield in-situ. The laser spot is focused on a single point of the sample and therefore an exact temperature measurement is not possible. With an infra-red camera, only an increase in the temperature is measured. With continuous laser irradiation, PL spectra are acquired every few seconds and the integrated PL yield is calculated by integration over the whole spectrum and is plotted in Figure 8a. First, in order to demonstrate the quality of the Sb$_2$Se$_3$/CdS stack in the 25°C reference state, a laser power of 10 mW is used resulting in a stable but relatively low PL yield (labeled: 1$^{st}$: 10 mW). Second, the laser power was set to 100 mW resulting in an instantaneous increase in the PL yield. During the duration for nearly 400 s, the PL yield increases with respect to time indicating an improving material quality with ongoing laser irradiation/annealing. Third, the sample is measured again at 10 mW (labeled 3$^{rd}$: 10 mW). Compared to the initial measurement at 10 mW (1$^{st}$), the PL yield increased by a factor of 3.3. In comparison, the sample annealed at 250 °C does not show an increased PL yield after going through the 100 mW laser irradiation/annealing.

Selected PL spectra of the reference sample are shown in Figure 8b at different times after the laser power was changed. The PL peak position under 100 mW excitation shifts to lower energies due to heating. Each spectrum in the time series is fitted with a pseudo-Voigt function (dash dotted lines) to determine the FWHM, which are plotted in Figure 8c. Initially, in the 1$^{st}$ stage at 10 mW, the FWHM for the reference sample is higher than for the *250 °C* sample. In the 3$^{rd}$ stage, i.e. after 100 mW irradiation, the FWHM values are lowered compared to the 1$^{st}$ stage and are for both samples rather similar. The fitted values for the FWHM confirm the trend observed for the PL yield (Figure 8a): Initially, the reference sample has a lower quality compared to the 250 °C sample, while after the 100 mW laser irradiation/annealing, the quality is similar.

Interestingly, the improvement in PL yield in the reference sample during the 100 mW laser irradiation/annealing is seen to rise fast initially, and then to start slowing down towards the end of the 400 s annealing period. It suggests that extending the previously used 300 s hot plate annealing would not bring any further benefit with respect to the improvement in the open circuit voltage.

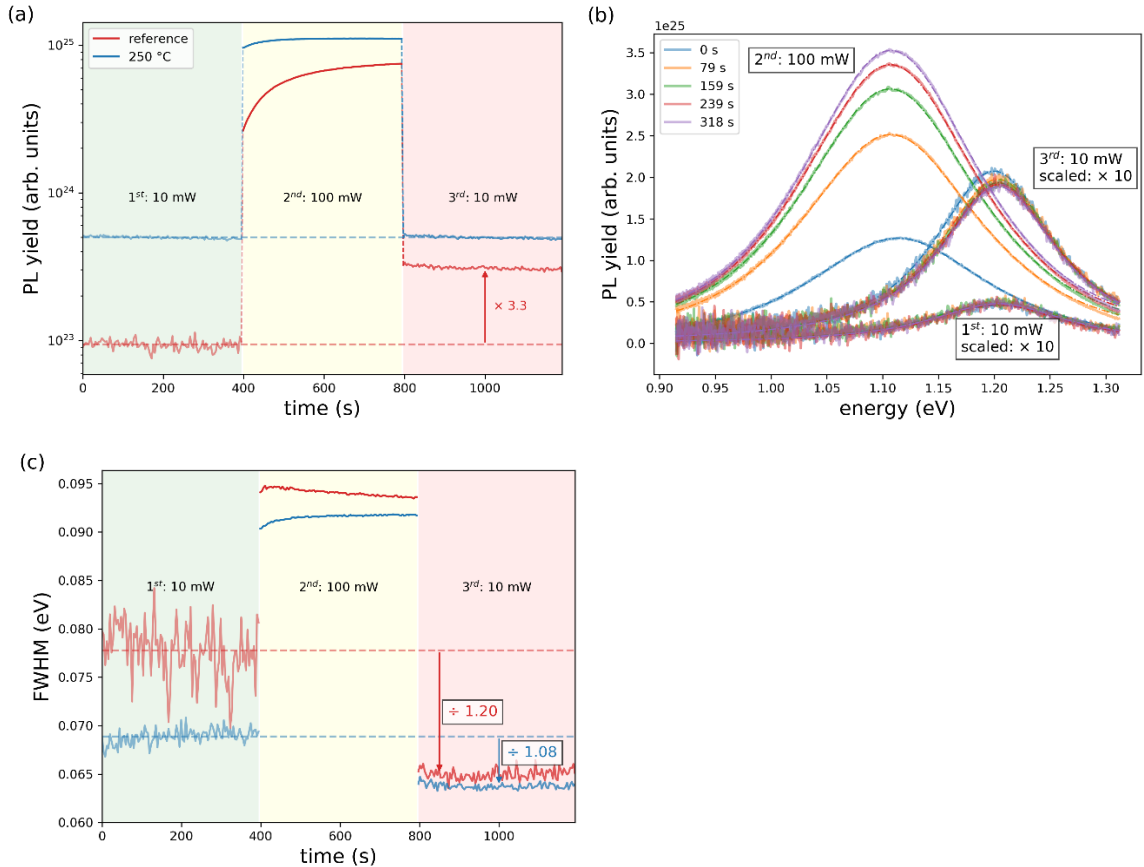

Figure 8 – a) Spectrally integrated PL yield with respect to time. The laser power was initially set to 10 mW (labeled 1st: 10 mW), then to 100 mW (labeled 2nd: 100 mW) and in the end again back to 10 mW (labeld 3rd: 10 mW). Dashed horizontal lines indicate the mean PL yield during the first 10 mW laser power irradiation. b) Time evolution of selected PL spectra acquired from the reference device. The legend indicates the time after the laser power was changed. The PL spectra are fitted using a pseudo-Voigt function (dash-dotted lines). c) Fitted FWHM values for the reference device (dark lines) and the device annealed at 250 °C (light lines).

## 3.5 Elemental depth profiling

Several reports exist, which indicate that the $Sb_2Se_3$/CdS interface is not stable and elemental diffusion and intermixing takes place at the interface [8, 13, 27-30]. In particular, for superstrate devices, it is assumed that an intermediate CdSe layer forms and has been identified to cause problems for the photocurrent collection [27, 31]. Elemental depth profiling of our substrate configuration $Sb_2Se_3$ absorber layers covered with a CdS buffer layer using SIMS could not detect a measurable compositional difference across the Sb$_2$Se$_3$/CdS interface between the reference and the 250 °C post-deposition annealed device (Figure 9a), which hints that no large CdSe layer forms within the error of the measurement. It is noted that the profiles shown in Figure 9 do not necessarily indicate inter-diffusion of Sb, Se, Cd and S but rather the consequence of the surface roughness that decreases the depth resolution of SIMS measurements. Indeed, as will be shown below, STEM-EDX measurements indicate the absence of elemental inter-diffusion.

It needs to be stressed that in superstrate configuration, the Sb$_2$Se$_3$ absorber is directly deposited onto the CdS buffer layer at higher temperatures than we used here to do the post-deposition annealing [27, 31], which likely leads to the formation of a CdSe layer for these growth conditions.

We observe a strong increase in alkali elements upon post-deposition annealing (Figure 9b) with highest counts at the Sb$_2$Se$_3$/CdS interface. These alkalis could alter the interface properties, as for instance observed in Cu(In,Ga)Se$_2$/CdS structures [32]. It remains to be investigated how these alkalis influence the interface recombination at the Sb$_2$Se$_3$/CdS interface, although alkali compounds such as NaSbS$_2$ are known to exist [33]. The oxygen at the interface did not notably change upon post-deposition annealing in air (Supplementary Figure 9b), which indicates that a SbO$_3$ phase does not form.

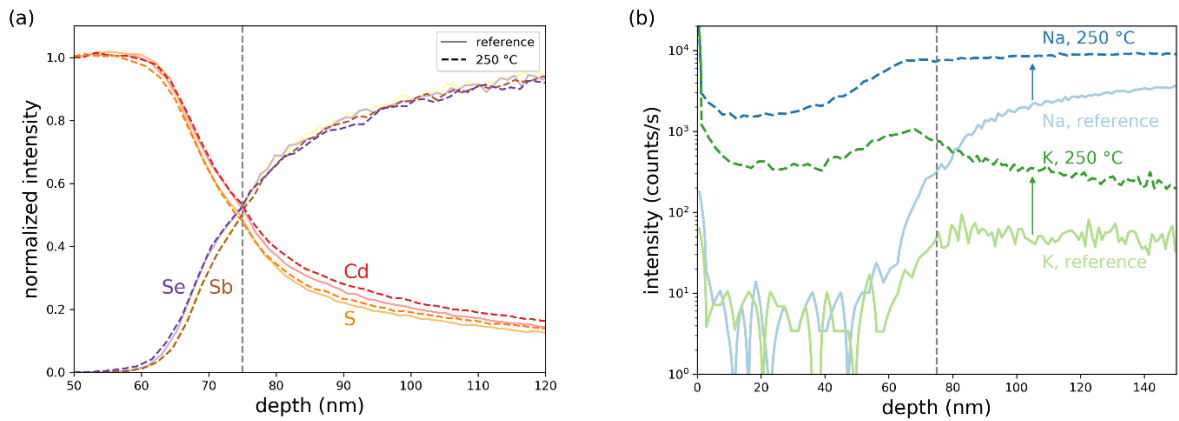

Figure 9 – a) Elemental gradients of Sb, Se, Cd and S across the Sb$_2$Se$_3$/CdS interface. The overlap of Se and Sb with Cd and S is the result of surface roughness. No differences between the reference and the 250 °C post-deposition annealed device is observed. b) Na and K alkali elements for the reference and the 250 °C post-deposition annealed device. Clearly, a significant increase of the alkali elements is observed upon post-deposition annealing with peak values at the Sb$_2$Se$_3$/CdS interface (gray dashed line). Note the different scale on the abscissa for a) and b). The combined plot of a) and b) is shown in Supplementary Figure 9a on a larger depth scale.

In order to investigate the Sb$_2$Se$_3$/CdS interface more in detail, STEM-EDX and HRTEM measurements are carried out. Figure 10a shows a high-angle annular dark field (HAADF) cross section and Figure 10b,c STEM-EDX elemental maps of Sb, Se, Cd, S and Mo for the device post-deposition annealed at 250 °C. The elemental maps display a CdS coating of 80 nm, the Sb$_2$Se$_3$ layer with thicknesses oscillating between 400 and 800 nm, and a MoSe$_2$ layer of 150 nm. Elemental profiles in Figure 10d of the region of interest selected in Figure 10a show no indication of elemental intermixing at the Sb$_2$Se$_3$/CdS interface, in agreement with the SIMS results. Figure 10f shows a HRTEM image of the Sb$_2$Se$_3$/CdS interface. A thin (approximately 2 nm) interface region is observed, which is a narrow overlapping region between adjacent areas as the result of the finite thickness of the TEM lamella and does not indicate another phase and/or material composition. The crystal structures at the interfaces are investigated by fast Fourier transforms (FFTs) within dedicated regions close and across the Sb$_2$Se$_3$/CdS interface (yellow – CdS, purple - Sb$_2$Se$_3$/CdS, and red Sb$_2$Se$_3$ region, respectively). All the FFTs correspond to those obtained from the Sb$_2$Se$_3$ and CdS crystal structures, which indicates that no secondary phase forms at the interface (see Supplementary Figure 10 for additional details).

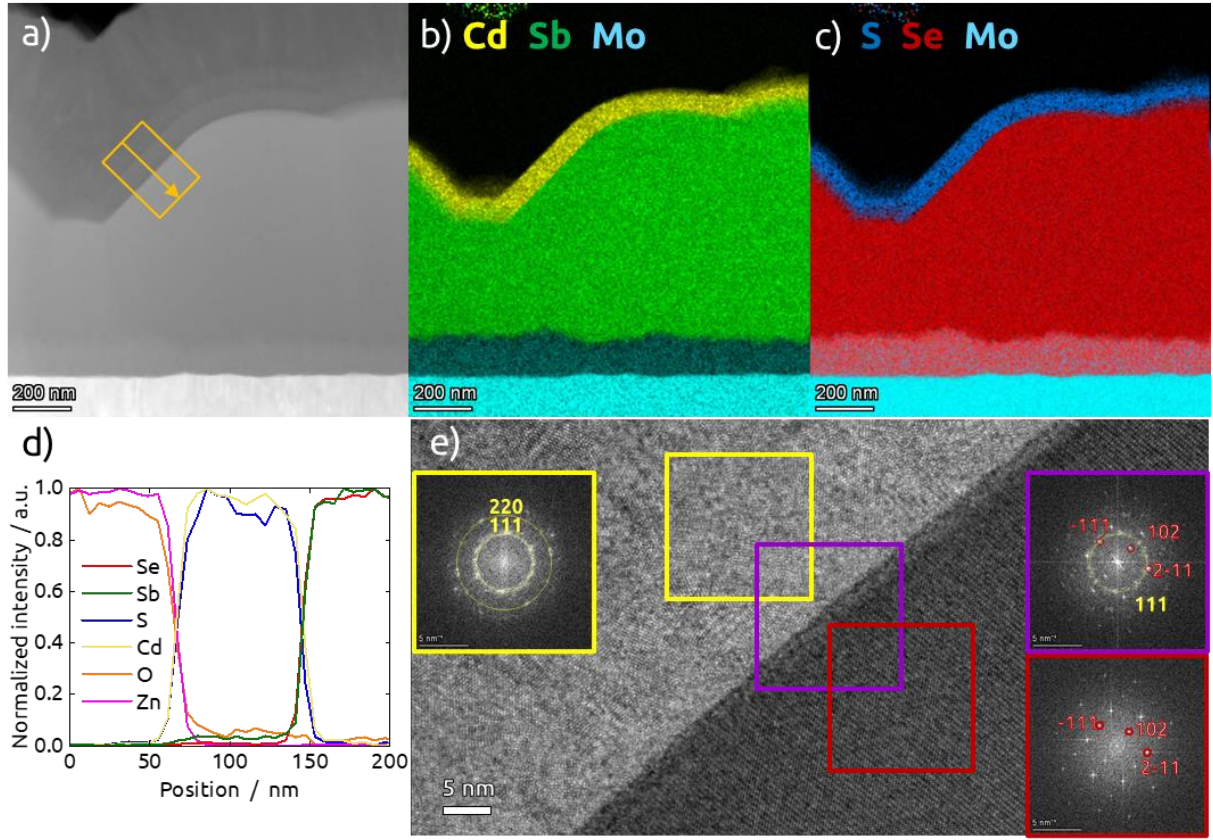

Figure 10 – TEM analysis of the solar cell device with a post-deposition annealing temperature of 250 °C. a) Cross-section STEM of the solar cell device; b,c) EDX mapping showing the presence of Cd, Sb, S, Se and Mo,; d) elemental profile of the selected region in a); e) HRTEM measurement of the $Sb_2Se_3$/CdS interface. FFTs (yellow – CdS, purple - $Sb_2Se_3$/CdS, and red $Sb_2Se_3$ region, respectively) are carried out to obtain the underlying crystal structure. No secondary phase is detected close to the interface and all diffraction characteristics can be explained by crystal structures of CdS and $Sb_2Se_3$.

### 3.6 ALD interfacial layers

As demonstrated in Figure 1 and from our own results here, interface recombination can impose a major bottleneck for the $V_{OC}$. Thus, to try to avoid this interface recombination, various interfacial ALD buffer layers are deposited between the $Sb_2Se_3$/CdS interface. It is reported that $TiO_2$ buffer layers improve device efficiency in substrate [8] and superstrate [27] configuration. However, until now, no reports exist of a $Sb_2S_3$ buffer layer on a $Sb_2Se_3$ absorber. Figure 11a shows illuminated and dark IV characteristics of the reference device, as well as devices with intermediate $TiO_2$ and $Sb_2S_3$ layers between the $Sb_2Se_3$/CdS interface. The interfacial layers improve all the photovoltaic parameters, as presented by the statistics in Figure 11c-f. $V_{OC}$ values increase from 340 mV for the reference device to 380 mV for the devices with interfacial layers. This improvement, similar to the post-deposition annealing, can be attributed to an increased activation energy of $J_0$ from 0.86 eV to 1.00 and 1.02 eV for the $TiO_2$ and $Sb_2S_3$ interfacial layer, respectively (Supplementary Figure 8). Very importantly, the interfacial ALD $Sb_2S_3$ and $TiO_2$ layers boost $J_{SC}$ values from 15 mAcm$^{-2}$ to average values of 24.0 and 22.3 mAcm$^{-2}$, respectively. The resulting efficiencies of the devices show a clear increase from 2.0 % for the reference solar cells, to 4.2 % and 3.7 % average efficiencies, being 4.6 and 4.2 % the highest efficiency values for $Sb_2S_3$ and $TiO_2$, respectively. The EQE spectra for the three different buffer layer configurations are shown in Figure 11b demonstrating an improved collection over the whole wavelength range. From the ratio of the EQEs (right ordinate), it is apparent that the collection improves in particular for longer wavelengths by up to 60 percent compared to the reference device.

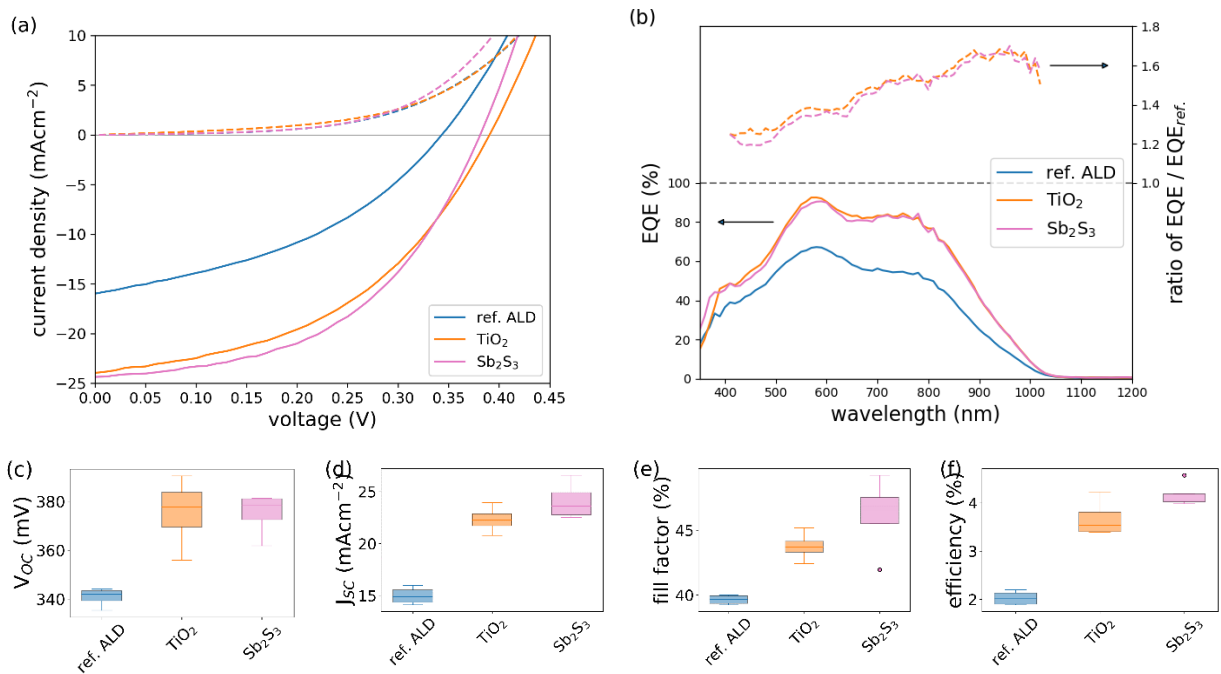

Figure 11 – Illuminated (solid lines) and dark (dashed lines) jV characteristics for a $Sb_2Se_3$/CdS reference device (ref. ALD) and with $TiO_2$ and $Sb_2S_3$ intermediate layers between the $Sb_2Se_3$/CdS interface. b) EQE curves (left ordinate) of the devices shown in a) indicating a strongly increased photo-current upon insertion of intermediate $TiO_2$ and $Sb_2S_3$ buffer layers. The increase (compared to the reference device) of the EQE is over the whole wavelength range, with increased contribution for higher wavelengths (right ordinate). c) – f) Statistics for $V_{OC}, J_{SC}, FF$, and efficiency, respectively.

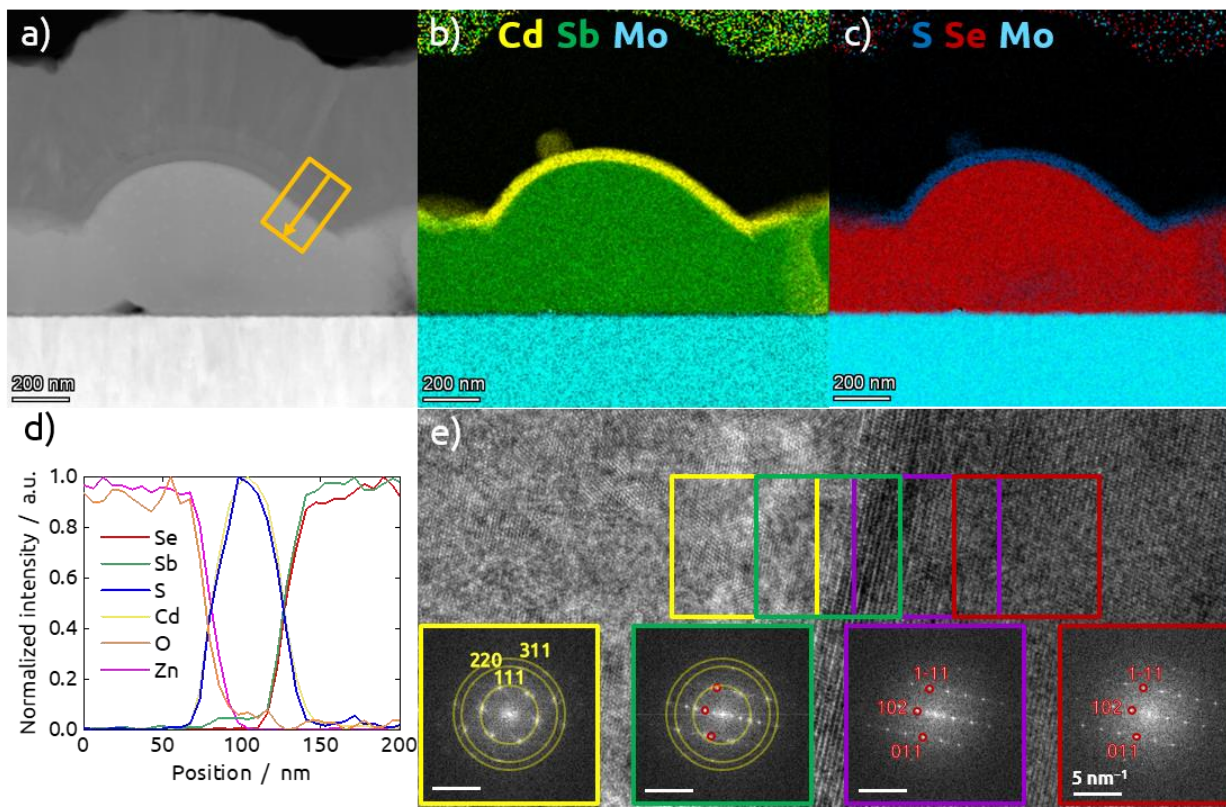

Figure 12: TEM analysis of solar cell devices based on $Sb_2Se_3$ layers selenized at 320 °C with an interfacial layer of $Sb_2S_3$ at the $Sb_2Se_3$/CdS interface. a) Cross-section STEM of the solar cell device. b,c) EDX mapping showing the presence of Cd, Sb, S, Se and Mo; d) Elemental profile of the selected region in a) ; e) HRTEM measurement of the $Sb_2Se_3$/ALD-$Sb_2S_3$/CdS interface. FFTs (yellow – CdS, green - $Sb_2Se_3$/CdS, purple - $Sb_2Se_3$ and red $Sb_2Se_3$ region, respectively) are carried out to obtain the underlying crystal structure. $Sb_2S_3$ interfacial layer is not detected. No secondary phase is detected close to the interface and all diffraction characteristics can be explained by crystal structures of CdS and $Sb_2Se_3$.

TEM cross-sections of a solar cell device with a $Sb_2S_3$ interfacial layer is presented in Figure 12a. EDX mappings in Figure 12b,c show the CdS and $Sb_2Se_3$ layers and the absence of $MoSe_2$ (due to selenization temperature of only 320 °C) at the $Sb_2Se_3$/Mo interface. Lower magnification TEM micrographs and elemental distribution maps expose the infiltration of CdS through pinholes in the $Sb_2Se_3$ (Supplementary Figure 11). Elemental profiles show that CdS and $Sb_2Se_3$ layers preserve a sharp interface (Figure 12d). The analysis of the FFTs taken along the $Sb_2Se_3$/$Sb_2S_3$/CdS interface from the HRTEM image further confirms the absence of secondary phases and the formation of single crystalline structures. In particular, no crystalline stibnite $Sb_2S_3$ is detected (Figure 12e and Supplementary Figure 15); possibly because an amorphous $Sb_2S_3$ is deposited and/or because the volume of $Sb_2S_3$ is below the detection limit. The HRTEM displays Moiré patterns at the interface due to the overlapping of two different crystal orientations attributed to the thickness of the lamella sample extracted from the device.

# 4 Discussion

## 4.1 Interface recombination

In the case of $Sb_2Se_3$ based solar cells, the activation energy of $J_0$ is generally reported below the bandgap independent of the device architecture (substrate or superstrate) and for different buffer and back contact layers [12, 22, 28, 29, 34-38]. An exception is reported in ref. [39], where an organic molecule is deposited between the window and the absorber layer. These activation energies point to dominating interface recombination due to a pinned Fermi level or a cliff-like band alignment [24]

The devices presented in Figure 7, in particular the devices after post-deposition annealing, show an activation energy of $J_0$ close or equal to the bandgap of the $Sb_2Se_3$ absorber layer, which is the highest value reported so far. These activation energies of $J_0$ directly influence the $V_{OC}$ via Eqn. (4) and can explain the increased values after post-deposition annealing up to 485 mV. It can be conjectured that the post-deposition annealing removes defects at the $Sb_2Se_3$/CdS interface. For instance, removal of acceptor-like defects close to the conduction band will result in (quasi) Fermi levels, which can move upon a bias voltage, i.e. Fermi levels which are not pinned, and thus an activation energy equals the $Sb_2Se_3$ bandgap is obtained [24]. Another possibility is the removal of defects upon post-deposition annealing, which are the origin of dipoles at the interface. Dipoles at the interface can induce a cliff-like band offset (minimum of the conduction band of the buffer below the conduction band minimum of the $Sb_2Se_3$ absorber) and thus a lowered interface bandgap and hence a reduced activation energy of $J_0$. The cases of a pinned Fermi level and a cliff-like band-alignment are simulated in the 1-dimensional solar cell simulator SCAPS [40]. In Supplementary Information section C, band-alignments, IV, QE and $V_{OC}(T)$ simulations are presented for the scenarios mentioned above. These simulations support the results presented here that the post-deposition annealing leads to increased $V_{OC}$ values due to an improved $Sb_2Se_3$ interface quality.

Recently, substrate $Sb_2Se_3$/CdS devices are fabricated with open circuit voltages above 500 meV [13]. These devices received a post-deposition annealing at 300 °C for 5 minutes after the CdS buffer layer deposition. The origin for the higher $V_{OC}$ might be the improved activation energy of $J_0$ and the accompanied reduced interface recombination, as demonstrated here.

Similar like the devices with a post-deposition annealing, also an interfacial ALD buffer layer improves the interface quality as evidenced by an increased activation energy of $J_0$ (Supplementary Figure 8), which in turn results in higher $V_{OC}$ values (Figure 11c). The increase activation energy of $J_0$ indicates that the improvement of Voc might be associated to reduction of detrimental recombination paths at the pn-junction by the interfacial layers. However, there is an additional beneficial effect of the interfacial layers since the photocurrent of the devices experience a drastic increase (Figure 11d) and is discussed in the section below.

In conclusion, as interface recombination constitutes a major recombination channel in substrate $Sb_2Se_3$/CdS based solar cells, the front surface area should be minimized to maximize $V_{OC}$.

## 4.2  Photo current losses and gains.

It was shown in section 3 that the material and interface quality improves upon post annealing due to

  i)    Increased PL yield and therefore a lowered non-radiative recombination current
  ii)   A decreased FWHM of the PL spectrum due to decreased disorder
  iii)  An increased activation energy of $J_0$ due to an improved $Sb_2Se_3$/CdS interface
  iv)   A lowered $J_0$ and hence an increased open circuit voltage

Yet, the short-circuit current density $J_{SC}$ decreases after post-deposition annealing. While this phenomenon is not fully understood, several scenarios can be ruled out.

- SIMS profiling and TEM analysis did not detect a measurable depth profile differences for Sb, Se, Cd, and S between the reference device and the device annealed at 250 °C (Figure 9 and Figure 10), unlike reported in [27]. In addition, from temperature dependent jV characteristics, no blocking of the photocurrent is observed for temperatures as low as 40 K (Supplementary Figure 12). Thus, even if a CdSe layer forms after post-deposition annealing, which cannot be resolved by SIMS or TEM, this layer does not act as a barrier for the photocurrent and thus cannot explain the reduced $J_{SC}$.

- Williams et al. observed a void formation in close-spaced sublimated Sb$_2$Se$_3$ films at the Sb$_2$Se$_3$/CdS interface and assigned those voids to a loss in $J_{SC}$ [31]. However, no voids were observed within the resolution limit of neither the SEM (Supplementary Figure 14) nor the TEM (Supplementary Figure 11).

Figure 4b show that photo-current losses are stronger for longer wavelengths. Thus, it might be possible that the post-deposition annealing reduces the collection efficiency of photo-generated charge carriers. Simulations shown in Supplementary Figure 19 show that a reduced electron mobility can have a strong effect on $J_{SC}$, where losses are strongest for long wavelengths. It was found previously that photo-carrier collection can depend on the grain orientation [41]. However, only a minor change in grain orientation is observed by XRD measurements (Supplementary Figure 13). It is possible though that the bulk mobility changed independent of the grain orientation and thus results in a decreased photo-current extraction.

Another way to have a reduced J$_{SC}$, as shown in Supplementary Figure 20, is to introduce a single acceptor-like defect at the Sb$_2$Se$_3$/CdS interface with varying distance to the Sb$_2$Se$_3$ valence band. The defect adds considerable negative charge at the Sb$_2$Se$_3$/CdS interface and therefore strongly influences the band bending in the Sb$_2$Se$_3$ absorber layer. With defect energies closer to the Sb$_2$Se$_3$ valence band, recombination is reduced, while collection is impeded at the same time. Thus, the model of an acceptor like interface defect close to the valence band can explain at the same time increased $V_{OC}$'s and reduced $J_{SC}$'s. It is thus possible that the same defect, in that case a reduction in density, is responsible for the increased $J_{SC}$ after surface modifications due to the ALD buffer layer deposition.

Ultrathin Sb$_2$S$_3$ and TiO$_2$ interfacial layer deposited via ALD at the Sb$_2$Se$_3$/CdS interface improves considerably the $J_{SC}$ of the devices(Figure 11d). EQE measurements demonstrate an improved charge carrier collection efficiency through the whole wavelength range (Figure 11b). The $J_{SC}$ values obtained by introducing Sb$_2$S$_3$ and TiO$_2$ interfacial layers are already close to the best $J_{SC}$ values reported literature for planar devices in substrate configuration (Figure 1).

The origin of the increase in photocarrier collection is currently still under investigation. Similar as for the devices with a post-deposition annealing, modifications (here a reduction) of a spike-like conduction band offset, and thus a reduced barrier for photo carriers, can be ruled out. This is supported by Supplementary Figure 12, which shows that the photo current has no strong temperature dependence.

Values of $J_{SC}$ around 30 mAcm$^{-2}$ or higher are only obtained for nano-ribbon Sb$_2$Se$_3$/CdS core-shell structures [8, 11, 12]. From the TEM analysis of the Sb$_2$Se$_3$/Sb$_2$S$_3$/CdS stack, i.e. a Sb$_2$S$_3$ interfacial ALD layer, it is observed that CdS intrudes into pinholes and voids at the back contact (Supplementary Figure 11). Thus, it is likely that the ALD layers passivate these surfaces similarly as the front surface [8, 14, 15]. The improved $J_{SC}$ can then be explained by increased charge-carrier collection via an increased front interface area similar as in the core-shell structures. However, as pointed out in section 4.1, the front surface constitutes a major recombination channel and thus questions the approach of core-shell structures. Ideally, with improving minority carrier lifetimes, i.e. improving the bulk quality of the Sb$_2$Se$_3$ absorber, also the carrier collection length increases. Thus, $J_{SC}$ will improve simultaneously, without the need of a large front surface area.

# 5 Conclusions

The limitations of the open circuit voltage for Sb$_2$Se$_3$/CdS substrate solar cells are investigated. It is shown that a post deposition annealing after the CdS buffer layer deposition can significantly improve the open circuit voltage due to an improved activation energy of $J_0$ resulting in a reduced non-radiative recombination. In particular, activation energies as high as the Sb$_2$Se$_3$ bandgap are demonstrated, which shows that major losses due to interface recombination (pinned Fermi level or a conduction band cliff) can be eliminated. SIMS and HRTEM measurements do not show the occurrence of another phase at the Sb$_2$Se$_3$/CdS interface. However, while $V_{OC}$ improves, $J_{SC}$ decreases and consequently limits the efficiency of the devices presented here. EQE measurements in combination with simulations indicate that the transport properties within the Sb$_2$Se$_3$ bulk might be the reason for the reduced $J_{SC}$. Similar to the post-deposition annealing, intermediate ALD buffer layers such das TiO$_2$ or Sb$_2$S$_3$ show an increased activation energy of $J_0$ as well and thus a reduced interface recombination. In contrast to the post-deposition annealing, these ALD interfacial layers improve $J_{SC}$ for our substrate devices and might pave the way to overcome the limitations observed in this study and thus for higher efficiencies in the future.

Importantly, as interface recombination acts as the major recombination channel in Sb$_2$Se$_3$ based substrate solar cells, the front interface area needs to be minimized and planar structures are the preferred architecture. Once the bulk Sb₂Se₃ properties are improved to yield a higher minority carrier lifetime, the photo current will improve naturally due to an increased diffusion length.

# 6 Acknowledgements

# 8 Appendix A

Table 1 – IV parameters for substrate and superstrate $Sb_2Se_3$ devices from literature.

| architecture | Efficiency (%) | $V_{OC}$ (V) | $J_{SC}$ (mAcm$^{-2}$) | ref |
|---|---|---|---|---|
| **superstrate** | 6.54 | 0.418 | 32.2 | [42] |
| | 5.72 | 0.371 | 26.01 | [43] |
| | 6.7 | 0.406 | 28.56 | [44] |
| | 3.7 | 0.335 | 24.4 | [45] |
| | 3.5 | 0.339 | 20.7 | [46] |
| | 5.76 | 0.383 | 26.3 | [47] |
| | 3.47 | 0.35 | 19.2 | [48] |
| | 6.29 | 0.394 | 27.2 | [37] |
| | 5.48 | 0.450 | 25.44 | [27] |
| | 3.73 | 0.329 | 22.8 | [49] |
| | 5.18 | 0.385 | 24.5 | |
| | 5.93 | 0.391 | 26.2 | [29] |
| | 4.81 | 0.357 | 28.1 | |
| | 5.14 | 0.356 | 25.28 | [50] |
| | 7.6 | 0.42 | 29.9 | [30] |
| | 6.63 | 0.42 | 32.5 | [51] |
| | 5.6 | 0.4 | 25.14 | [41] |
| | 7.5 | 0.432 | 27.6 | [52] |
| | 7.50 | 0.423 | 30.8 | [53] |
| | 6.5 | 0.427 | 25.5 | [54] |
| | 6.84 | 0.421 | 28.4 | [38] |
| | 6.32 | 0.413 | 28.1 | [55] |
| | 7.3 | 0.434 | 31.7 | [56] |
| | 5.23 | 0.40 | 25.1 | [57] |
| | 5.6 | 0.446 | 25.99 | [58] |
| | 6.24 | 0.38 | 28.1 | [59] |
| | 5.4 | 0.36 | 29.0 | [60] |
| | 7.45 | 0.413 | 28.9 | [61] |
| | 7.02 | 0.414 | 29.2 | [62] |
| | 5.89 | 0.382 | 29.6 | [63] |
| | 5.03 | 0.40 | 22.07 | [39] |
| | 6.6 | 0.394 | 29.5 | [64] |
| | 7.15 | 0.426 | 29.3 | [65] |
| | 6.33 | 0.41 | 27.45 | [66] |
| | 7.29 | 0.41 | 30.94 | [67] |
| | 7.89 | 0.449 | 28.3 | [68] |
| **substrate** | 1.47 | 0.407 | 12.11 | [34] |
| | 4.25 | 0.427 | 17.11 | [69] |
| | 9.2 | 0.40 | 32.58 | [8] |
| | 6.71 | 0.403 | 25.69 | [70] |

| | 2.1 | 0.354 | 17.8 | [35] |
| --- | --- | --- | --- | --- |
| | 6.15 | 0.455 | 22.75 | [36] |
| | 6.06 | 0.494 | 25.91 | [28] |
| | 4.3 | 0.386 | 19.8 | [71] |
| | 6.84 | 0.504 | 24.91 | [13] |
| | 5.7 | 0.422 | 21.9 | [22] |
| | 6.43 | 0.446 | 26.4 | [72] |
| | 9.19 | 0.461 | 29.92 | [12] |
| | 4.22 | 0.454 | 18.4 | [73] |
| | 8.46 | 0.402 | 30.68 | [11] |